\documentclass[aps,twocolumn]{revtex4}

\usepackage{graphicx}
\begin{document}
\title{Understanding the saturation of proton-driven Weibel instabilities 
and implications for astrophysics}
\author{Chuang Ren$^{\dag \ddag \S}$,  Eric G. Blackman$^{ \ddag \S}$, and Wen-fai Fong*\\}
\address{\\{\dag Department of Mechanical Engineering}
\\{\ddag Department of Physics and Astronomy}
\\{\S Laboratory for Laser Energetics}
\\{University of Rochester, Rochester, NY 14627}
\\ {*}Massachusetts Institute of Technology, Cambridge, MA 02139\\}
\date{\today}                                           
\begin{abstract}
{The linear growth rate and saturation level of magnetic fields for Weibel instabilities driven 
by  ion temperature anisotropy, defined as $\alpha=(T_\perp/T_\parallel)-1$ where $T_\perp$ and $T_\parallel$ are ion temperatures perpendicular and parallel to the wave vector, are derived in the small $\alpha$-limit. It is shown that the ratio of the saturated magnetic energy to the initial ion energy scales as the fourth power of the electron to ion mass ratio, $m/M$, for an initially unmagnetized plasma with $\alpha\leq M/m$.  Particle-in-cell simulations confirm the mass scaling and also show that the electron energy gain is of the same order of magnitude as the magnetic field energy.
This implies that the Weibel instabilities cannot provide a faster-than-Coulomb collisionless 
mechanism to equilibrate ion-electron plasmas with ions initially much hotter 
than electrons, a key component in low-luminosity astrophysical accretion flows. 
The results here also show that the large $\alpha$-limit formulae used in the study of magnetic field generation in collisionless shocks are only valid if $\alpha\geq M/m$.}
\end{abstract}
\maketitle

\narrowtext
\section{Introduction}

\subsection{Weibel instabilities and astrophysical shocks}

{
Weibel instabilities, whereby an initially anisotropic plasma distribution
relaxes to convert the free energy in the anisotropy into magnetic fields
e.g. \cite{w59,davidson72,ml99} 
have received much recent attention in the study of astrophysical
plasmas, particularly in the context of influence on the observed 
emission associated with collisionless astrophysical shocks.
These include shocks of the crab pulsar wind \cite{yang94}, gamma-ray bursts (GRB)\cite{ml99}, and 
between or within  galaxy clusters \cite{msk06,markevitch06}.  The former two are relativistic.
The roles that the Weibel instabilities may play 
via their ability to grow magnetic fields  are (i) to 
magnetically trap particles near the shock, facilitating the 
shock formation itself; (ii) to 
account for synchrotron or jitter radiation that would otherwise
be absent and hard to explain without a source of field growth; 
(iii)  to suppress  thermal conductivity, as needed to explain
observed temperature inhomogeneities \cite{msk06} in the ion-electron
plasmas of Galaxy clusters; and
(iv) possibly producing a faster-than Coulomb coupling or
electron acceleration in the post-shock regions near shocks
within or between Galaxy clusters. This is important for the estimate of Galaxy cluster masses and therefore fractional dark matter mass fraction estimates \cite{markevitch06}.
The latter are inferred from measured electron 
temperatures, the assumption that proton temperatures are equal to the
X-ray inferred electron temperatures, and virial equilibrium.

 The plasma compositions for the astrophysical applications 
range from  ion-electron to positron-
electron pair plasmas, to a mixture between the two.
In shocks, the systematic velocity anisotropy is typically assumed to be strong, and is produced by 
fast particle streaming in the shock propagation direction compared to 
the directon parallel to the shock.  Current fluctuations
parallel to  the direction of  shock propagation
 lead to current filaments that amplify the magnetic field
in the shock plane, which then  further enhances the current filaments. }

\subsection{Weibel instabilities and two-temperature plasmas}

{We introduce the question of whether the Weibel instability may 
also facilitate a  faster-than-Coulomb coupling between ions and electrons
in two-temperature plasmas even unshocked.  
The free energy in anisotropic
need not be the result of particle streaming perpendicular to  shocks, but 
would also result from anisotropy in the local distribution of 
random particle velocities of the  species supplying the free energy
in the absence of shocks. The study of the Weibel instability
without shocks also has a long history \cite{davidson72} but has not been studied for plasmas containing ions and electrons at  very different average eneriges,the motivation for which we now describe further.}

Left alone, a two-temperature ion-electron plasma, with the ions 
at a higher temperature than the electrons,  will eventually 
approach  
an equilibrium in which the two species have the same temperature. 
Coulomb collisions provide a minimum equilibration rate, but
whether a faster-than-Coulomb collisionless equilibration mechanism
exists\cite{begelman,blackman99,quataert99} for specific particle distributions and external conditions is both a fundamental question in plasma physics and important for understading 
aspects of collisionless astrophysical phenomena such as radiatively 
inefficient accretion flows (RIAF) (e.g. Ref.\cite{quataert01}).

The survival of a two-temperature plasma for a Coloumb equilibration
time scale is important for  geometrically thick RIAFs 
of which 
advection dominated accretion flows (ADAFs) have been the archetype
\cite{ichimaru77,rees82,narayan95,narayan98}. 
In contrast to standard geometrically thin accretion discs
(e.g. Ref.\cite{shakura73}), 
which can comfortably allow 10\% of the binding energy of the
accreting material to be converted into observable
radiation around black holes, two-temperature accretion disks
can in principle reduce the photon luminosity by
many orders of magnitude for the same accretion rate.
Indeed,  the measured luminosities from the central engines of
some nearby elliptical galaxies seem to be  3 to 5
orders of magnitude smaller than expected for their estimated
accretion rates.
Similar quiescient phases of accretion are also
observed in some states of X-ray binaries.  There is also X-ray evidence that in some supernova remnants the ion temperature is much higher than the electron temperature\cite{vink05}.

In the simplest ADAF model, the reduced luminosity 
 results from three key features: (1) energy dissipated by viscosity 
is assumed to heat the protons, (2) only electrons radiate efficiently, 
and (3) if the only means of
energy transfer between electrons and protons is Coulomb collisions,
then a sufficiently collisonless plasma can be accreted onto
the central object  before the electrons receive enough energy to
radiate the energy dissipated by  the accretion. 
The gravitational binding energy of the accreting material is then 
retained as internal energy of the hot protons, and can be quiescently 
advected through the event horizon of a black hole.
Since the associated internal energy per ion in an ADAF is of order $m_pc^2 \sim$ 1GeV, the 
electrons are required to stay at a temperature much lower than the GeV ion temperature to ensure a low luminosity for a given accretion rate. 
ADAFs are thick discs because the thermal energy of the ions puffs up the
disk. 
More general models of two temperature accretion 
involving  outflows and convective feedback are also popular\cite{qua_nar99,bla_beg99,naretal02}.

The assumption that heat transfer from the ions to the electrons occurs
only by Coloumb collisions is necessary to ensure that 
a negligible fraction of the dissipated thermal energy 
is transferred to the electrons during the accretion infall time. 
This assumption has yet to be validated or invalidated (e.g. Ref.\cite{pariev05}).
Doing so requires understanding the subtle plasma physics of the
interactions among the ions, electrons,  and electromagnetic
fields. A systematic way to make  progress 
is to test this assumption under a range of  specific conditions. 
If the assumption is found to be violated for particular circumstances, 
then the associated instability and faster-than-Coulomb coupling 
would  limit the applicability of  two-temperature accretion paradigm. 
If the assumption is not violated for a particular
set of conditions, 
this would not prove the viablity of two-temperature plasmas 
in  all circumstances and more cirumstances would need to be tested.
While the latter issue has discouraged some people from working on this 
problem, we think that a series of rigorous tests using different initial
and boundary conditions is needed:  Were 
the assumption to emerge as consistently valid for 
a range of different initial conditions, this would
gradually solidify the robustness of the assumption
for an increasingly wider range of viable conditions,
and justify two-temperature accretion flows for those
regimes tested.

In this spirit,  we study one test of the Coulomb equilibration  
assumption here, namely, whether ion-electron energy transfer can be sped up via Weibel instabilities driven by an ion distribution anisotropy. For a real disk, the 
microphysics by which gravitational potential energy is
 converted to ion thermal energy is not fully understood, but 
presumably  involves turbulence generated by a
 collisionless version \cite{sharma03} 
of a magneto-rotatioanl instability \cite{balbus98}. MHD turbulence is believed to anisotropically cascade \cite{goldreich97} and it is likely in a collisionless system,
the proton temperature may also incur anisotropies on small scales,
transiently induced by the cascading turbulence. Similar to an anisotropic electron distribution, an anisotropic ion distribution  can also drive a  
Weibel instability \cite{w59,davidson72} which amplifies fluctuating magnetic fields. How fast and how much energy the electromagnetic fields and the electrons then gain from the isotropizing protons is the subject of our present work.

\subsection{Growth and Saturation of the Weibel Instability}

{Since  electrons gain energy from  electromagnetic fields generated  the Weibel instability, the electron 
energy gain is closely related to the energy level of the magnetic field, the dominant electromagnetic field at saturation. 
The strength of the saturated magnetic field level from the Weibel instabilities is therefore important for all of the astrophysical applications discussed above:  two temperature accretion plasmas as well as the origin of magnetic fields in Gamma-ray bursts\cite{ml99,wa04} and galaxy clusters \cite{msk06}.
The saturation mechanism is also a fundamental plasma physics question.

If energy transfered to electrons  were a significant fraction of the initial ion energy during an instability growth time for the two temperature
plasmas discussed in the previous section, 
this transfer would be a much more efficient process than Coulomb collisions for temperature equilibration. The is because   the ratio of maximum growth rate of the ion-driven Weibel instability $\gamma_{max}$ (see Eq.\ref{gmax} below) to collisional ion-electron equilibration rate $\nu_{eI}$ is 
\begin{equation}
\label{governu}
{\gamma_{max}\over\nu_{eI}}=7.2\alpha^{3/2}{n\lambda_{D}^3\over ln\Lambda}\sqrt{{T_e\over Mc^2}}\left(1+{T_{I}m\over T_{e}M}\right)^{3/2}.
\end{equation}
Here $\alpha \ge 0$ is a measure of ion temperature anisotropy (defined more precisely in Sec.\ref{theory}), $n$ the electron density, $\lambda_{D}$ the Debye length, $ln\Lambda$ the Coulomb logarithm, $T_e$ and $T_I$ the electron and ion temperatures (multiplied by the Boltzmann constant, as throughout the paper), respectively, $m$ and $M$ the electron and ion masses, respectively, and $c$ the velocity of light. For a proton-electron plasma with $n=10^{12} cm^{-3}, T_I=10^{12}K, \alpha=1,$ and $T_e=T_Im/M, {\gamma_{max}/\nu_{eI}}\sim 3\times10^7$. This ratio is large, but  the question is what fraction of the ion energy will be transfered to the electrons when the instability saturates?

There have been  two saturation mechanisms 
discussed in the literature: (i) the deviation from the equilibrium orbits of the particles with an anisotropic distribution \cite{davidson72,yang94,wa04} and (ii) the magnetization of the ions \cite{ml99} or electrons \cite{msk06}. These two mechanisms gave the same magnetic field level in the electron-driven case (for either ion-electron or pair plasmas) 
but differ in the ion-driven case \cite{wa04}. In the present paper 
we will  show and reaffirm that  the orbit deviation is the correct magnetic saturation mechanism for the Weibel instabilities. 

We will derive the growth rate and magnetic saturation level in the small anisotropy limit in Sec.\ref{theory} which shows that the fraction of the ion energy transfered to the magnetic field via the Weibel instabilities scales as $(m/M)^4$ for a single mode. Connection will also be made to previous results in the large anisotropy limit. Particle-in-Cell (PIC) simulations in Sec.\ref{sim} with reduced ion mass will confirm this mass scaling and also will show that the electron energy gain is of the same order of magnitude as the magnetic field energy.
For an initially unmagnetized plasma, this implies that the Weibel instabilities are  insignificant to the ion-electron temperature equilibration process.
However, in Sec.\ref{dis} we discuss how 
the transfered energy fraction via the Weibel instabilities may greatly
increase for an initially more strongly magnetized plasma. 
We conclude  in Sec. \ref{dis}.}

\section{Caluclating 
growth and saturation of proton-driven Weibel instabilities}
\label{theory}
\subsection{The growth rate}
\label{growth}
The linear growth rate $\gamma$ of the Weibel instability has been derived for various anisotropic electron or ion distributions \cite{w59,davidson72}.
For the ion driven case,  Ref. \cite{davidson72} derived $\gamma$ in the large anisotropy (growth rate) and low plasma temperature limit for the ``ion-pinch'' case, where the electron distribution is an isotropic Maxwellian and the ion distribution consists of two bi-Maxwellian counter-streams \cite{davidson72}. 
Here we also study instability driven by the ions, but we do not
invoke any  ion bulk streaming, and instead allow the instability to develop
from  ion temperature anisotropy alone. Also in contrast to \cite{davidson72}, 
we focus on the 
the small anisotropy (growth rate), high ion  temperature limit. 

We assume that the electrons have an initially isotropic electron distribution given by 
\begin{equation}
\label{fe0}
f_{e0}({\bf v}, {\bf x})={n\over(2\pi T_e/m)^{3/2}}\exp\left(-{mv^{2}\over2T_e}\right), 
\end{equation}
while the ions take on a bi-Maxwellian distribution, 
\begin{equation}
\label{fi0}
f_{I0}({\bf v}, {\bf x})={n\over(2\pi/M)^{3/2}T_{I\perp}\sqrt{T_{Ix}}}\exp\left(-{Mv_{\perp}^{2}\over2T_{I\perp}}\right)\exp\left(-{Mv_x^2\over2T_{Ix}}\right).
\end{equation} Here, $n$ is the density of the electrons or the ions (singly charged ions are assumed for simplicity), $T_e$ the electron temperature, 
$v_\perp^2=v_y^2+v_z^2$,  and $T_{Ix}$ and $T_{I\perp}=T_{Iy}=T_{Iz}$ are the ion temperatures in the $x$-direction and in the directions perpendicular to $x$ respectively.
To study the linear stability of this system against the Weibel instability, we assume that the mode wave vector is in the $x$-direction, ${\bf k}=k {\hat x}$, and all perturbations have spatial and temporal dependence of $\exp(ikx-i\omega t)$. For simplicity, we do not consider any coupling between transverse and longitudinal modes\cite{ren04,tzouf06} and assume that the modes are transverse, ${\bf E}=E_y{\hat y}$ and ${\bf B}=B_z{\hat z}$. 

We follow the procedure of Ref. \cite{davidson72} in linearizing
the  Vlasov-Maxwell system of equations, which leads to
the following dispersion relation,
\begin{equation}
\label{full_disp}
{k^2c^2\over\omega^2}+{\omega_{pe}^2\over\omega^2}[1+{1\over2}Z'(\xi_e)]+{\omega_{pI}^2\over\omega^2}[1+{1\over2}Z'(\xi_I)]+{\omega_{pI}^2\over2\omega^2}\alpha Z'(\xi_I)-1=0.
\end{equation}
Here  $\omega_{pe}^2\equiv4\pi q^2 n/m$ and $\omega_{pI}^2\equiv4\pi q^2 n/M$ are the plasma frequency of each species, $\xi_e\equiv\omega/(k\sqrt{2T_e/m})$ and $\xi_I\equiv\omega/(k\sqrt{2T_{Ix}/M})$, $Z'$ is the derivative of plasma dispersion function $Z(\xi)\equiv \pi^{-1/2}\int^{\infty}_{-\infty} {exp(-\lambda^2)/( \lambda-\xi)}d\lambda$ \cite{fried:z}, and 
$\alpha\equiv T_{I\perp}/T_{Ix}-1$ represents the ion temperature anisotropy. Eq.\ref{full_disp} can also be directly derived from Eq.(37) in Ref.\cite{davidson72} by removing bulk streaming from both species and assuming an isotropic electron distribution. 
Verified {\em a posteriori}, the growth rate for the ion-driven Weibel instability for small to moderate $\alpha$ is small, $|\omega|/kc\sim m/M$. Therefore, we take the small argument limit, $\xi<<1$, for $Z'(\xi), Z'(\xi)\approx -2-2i\sqrt{\pi}\xi$, to further simplify the dispersion relation Eq.\ref{full_disp}. In this small growth rate limit,  the displacement current term (the constant 1 on the left-hand side) can also be neglected. This leads to
\begin{equation}
\label{rate}
\omega=i\sqrt{2\over\pi}k\sqrt{{T_{e}\over m}}{m\over M}\left(\alpha-{k^2c^2\over\omega_{pI}^2}\right)\left(1+\beta\right)^{-1},
\end{equation}
where $\beta\equiv\sqrt{mT_e\over MT_{Ix}}(1+\alpha)$ is usually much less than 1 when $T_e\leq T_{Ix}$ and $\alpha \sqrt{m/M}<<1$.
Eq. (\ref{rate}) shows that the mode is unstable when $\alpha>0$ and 
\begin{equation}
k^2<k_0^2\equiv\omega_{pI}^2\alpha/c^2.
\end{equation}
In fact, this expression for the range of unstable $k$'s is exact and can be obtained from Eq.\ref{full_disp} without using the small $\xi$ approximation of $Z$. The maximum growth rate is reached at $k^*=k_0/\sqrt{3}$, and is given by
\begin{equation}
\label{gmax}
\gamma_{max}=\sqrt{8\over27\pi}\omega_{pI}\sqrt{{T_{Ix}\over Mc^2}}\alpha^{3/2}{m\over M}\sqrt{T_e/m\over T_{Ix}/M}(1+\beta)^{-1}.
\end{equation}
 Eq. (\ref{gmax}) (with $\beta\approx 0$) was used to derive Eq. (\ref{governu}).
 
{ From Eq. (\ref{gmax}) it is clear that the $\gamma_{max}$ is smaller than $\omega_{pI}$ by a factor of $m/M$, when $T_e/m\sim T_{Ix}/M$, in this small $\xi_I$ limit, valid for small to moderate $\alpha$. (For $T_e\sim T_{Ix}$, this factor is $\sqrt{m/M}$.) Since the linear growth rate determines the saturation level of the magnetic field (see Sec.\ref{saturation}), it is interesting to see how large  $\gamma$ can become in the large $\xi_I$ limit, valid for large $\alpha$. Since $Z'(\xi)\approx1/\xi^2$ when $\xi>>1$, we obtain from Eq.\ref{full_disp}\cite{davidson72},
\begin{equation}
\label{large_rate}
\gamma=\omega_{pI}\sqrt{{T_{Ix}\over Mc^2}}{\sqrt{1+\alpha}(kc/\omega_{pI})\over\sqrt{1+(kc/\omega_{pI})^2+M/m}},
\end{equation}
which is valid for $k<<k_0$. Using Eq.\ref{large_rate} we can find that it takes a significantly large anisotropy, $\alpha>>\sqrt{2}M/m$, to achieve $\xi_I>>1$. For large $k$, i.e., $kc/\omega_{pI}>>\sqrt{M/m}$, $\gamma$ approaches the maximum value of $\omega_{pI} \sqrt{T_{I\perp}/Mc^2}$, which is still less than $\omega_{pI}$ for any non-relativistic $T_{I\perp}$. A relativistic calculation using a water-bag distribution found that $\gamma\sim\omega_{pI}$\cite{wa04}. Our discussion here indicates that this is only possible for relativistic temperatures and extremely high $\alpha$'s. For $\alpha\sim M/m$, Eq.\ref{gmax} scales the same as  the maximum growth rate from Eq.\ref{large_rate}, 
\begin{equation}
\gamma\sim\omega_{pI}\sqrt{{T_{Ix}\over Mc^2}}\sqrt{M\over m}\sim\omega_{pI} \sqrt{T_{I\perp}\over Mc^2}
\end{equation}
It is thus reasonable to infer that Eq.\ref{gmax} has the correct scaling for $\alpha$ up to $M/m$.}

\subsection{Saturation}
\label{saturation}
 The PIC simulations of Ref.\cite{davidson72} showed that  the linear growth for a particular $k$-mode saturates when the magnetic field amplifies to $B_s$ defined in the following expression
\begin{equation}
\label{sat}
\gamma_k\sim\sqrt{kV_\perp qB_s/Mc}.
\end{equation}
Here, $\gamma_k$ is the linear growth rate for the mode $k$ and $V_\perp$ and $M$ are the characteristic perpendicular (to $k$) velocity and mass of the instability-driving species (the ions in our case). Reference \cite{davidson72} identifies the right-hand side of Eq.\ref{sat} as the magnetic bounce frequency and concluded that the saturation results from particle-trapping. Detailed analytical and numerical analysis in Ref.\cite{yang94} showed that Eq.\ref{sat} is equivalent to the condition $k\delta x\sim1$ where $\delta x$ is the perturbation to the ion orbits in the $k$-direction due to the electromagnetic fields of the instability. In another word, the magnetic field stops growing when the deviation of ion orbits is comparable to the instability wavelength. Here we give still another explanation of Eq.\ref{sat} as a condition that the current perturbation from the particle bouncing and trapping reaches the maximum possible value in one $e$-folding time.

The Weibel instability is driven by the currents created by and located near the nulls of the perturbed magnetic field. This can be seen most simply by assuming that initially all particles move only in the $\pm y$ -direction 
with equal numbers moving at velocities $-V_y$ and $+V_y$, resulting in zero $j_y$ everywhere. When there is a perturbed magnetic field $B_z$ with a $k_x$,  because of the sign change of  $B_z$ near a certain null, the particles moving in one direction are trapped (in the $x$-direction) near the null and the particles moving in the opposite direction are deflected away from the null, creating a net current.  Near the neighboring null, both the trapped and deflected particles move in reversed directions, resulting a net current of an opposite sign. It is these alternating currents, represented by the $\alpha$-term in Eq.\ref{full_disp}, that drive the instability. The net current density $j_y$ created depends on how effectively the current of the opposite sign can
be reduced by transverse deflection of particles away from the region
 of interest. The magnitude of the current can therefore 
be estimated from the deflected particle flux by
\begin{equation}
{\partial j_y\over\partial t}\approx{\partial\over \partial x}(j_M v_x),
\label{8}
\end{equation}
where $j_M=(1/2)n|qV_y|$ represents the maximum current density 
achievable, which would arise if all particles of one velocity sign
leave the null region. 

Due to magnetic deflection,  $\partial v_x/\partial t\approx \omega_c V_y$ where $\omega_c=qB_z/Mc$ is the cyclotron frequency. Since $\partial B_z/\partial x\approx kB_s$ near the null if $B_s$ is the magnetic field amplitude, we obtain
\begin{equation}
\label{jy}
{\partial^2 j_y\over\partial t^2}\approx j_M {qkB_sV_y\over Mc}.
\end{equation}
Comparing Eq. (\ref{sat}) and Eq. (\ref{jy}), we can see that $B_s$ is the field strength such that  the current growth in one $e$-folding time reaches  $\sim j_M$, after which the current can no longer increase and the mode saturates. This saturation mechanism also shows that  $q, V_\perp$, and $M$, in Eq. (\ref{sat}) should be that of the locally anisotropic species (the ions in our case) since the net current creation from the trapping and bouncing would be zero for an isotropic distribution \cite{fried59}. 

Combining Eq.(\ref{sat}) with Eq.(\ref{rate}) or Eq.(\ref{large_rate}) we can derive the saturated level for the magnetic field $B_s$ for $\alpha$ small or large. In the small $\alpha$ limit, Eq.(\ref{sat}) and Eq.(\ref{rate}) show that the largest $B_s$ occurs at a mode number of $k_m=k_0/\sqrt{5}$, which is smaller than that of the fastest-growing mode, $k^*$. The ratio of magnetic energy in this mode at saturation to the initial proton kinetic energy gives
\begin{eqnarray}
\label{ebk}
{E_B\over E_{kp}}&\equiv&{B_{s}^2/8\pi\over nT_{Ix}(3+2\alpha)}\nonumber\\
&\sim&1.5\times 10^{-2}{\alpha^5\over 3+2\alpha}\left({m\over M}\right)^4 \left({T_e/m\over T_{Ix}/M}\right)^2(1+\beta)^{-4}.
\end{eqnarray}
Because of the mass ratio factor, in a single mode $E_B/ E_{kp}\sim2.6\times 10^{-16}$ for an electron-proton plasma with $\alpha=1$ and $T_e/m=T_Ix/M$. This would also approximates the total magnetic field energy if the spectrum at saturation is dominated by this mode.  Because the magnetic energy is only a small fraction of the proton kinetic energy and
because the only way for electrons to gain energy is through the
magnetic energy,  only a negligible fraction of the proton energy can 
be expected to be transferred to electrons  when the Weibel instability saturates.  
The PIC simulations in Sec.\ref{sim} will indeed show that the electron energy gain is  on the same order of magnitude as the magnetic field energy gain. 
The upshot is that for an initially unmagnetized ion-electron plasma with moderate anisotropy, 
the Weibel instability cannot generate electromagnetic fields
of sufficient strength to energetically couple them  to 
the protons. 

{For $\alpha\sim M/m$, Eq.(\ref{ebk}) gives $E_B/ E_{kp}\sim 4.7\times10^{-4}\sim O(m/M)$. To obtain more accurate results for large $\alpha$, Eq.(\ref{large_rate}) and Eq.(\ref{sat}) can be used to obtain
\begin{eqnarray}
\label{ebk2}
{E_B\over E_{kp}}\approx {1\over4}{(kc/\omega_{pI})^2\over[1+(kc/\omega_{pI})^2+M/m]^2}.
\end{eqnarray}
Eq.(\ref{ebk2}) shows that in the large $\alpha$ limit the largest $B_s$ occurs at a mode number of $k^\dag=(\omega_{pI}/c)\sqrt{M/m}=\omega_{pe}/c$, 
$E_B/E_{kp}\sim m/(16M)$. This result agrees with that from the calculation using the water-bag distribution\cite{wa04}, $E_B/E_{kp}\sim m/(6M)$, within a numerical factor.

In the literature,  another saturation criterion was proposed\cite{ml99,msk06}, 
$k\rho\sim1$, where $\rho$ is the Larmor radius of one of the species in the plasma. For the Weibel instabilities driven by the electrons in a stationary ion back ground, this gives similar scaling for $B_s$ as Eq.\ref{sat} if $k=\omega_{pe}/c$ and $\rho=\rho_e$, the electron Larmor radius,  are used. However, for the proton-driven case here, Ref.\cite{wa04} pointed out that this criterion with $k=\omega_{pI}/c$ and $\rho=\rho_I$, the ion Larmor radius, used in Ref.\cite{ml99} gave a larger $B_s$ than Eq.(\ref{sat}). Recently, Ref.\cite{msk06} claimed that this criterion gave the correct scaling, $E_B/E_{kp}\sim m/M$, if $k=\omega_{pI}/c$ and $\rho=\rho_e$ were used. The analysis here shows that the largest $B_s$ occurs at $k=\omega_{pe}/c$, not $k=\omega_{pI}/c$. At $k=\omega_{pI}/c$, Eq.\ref{ebk2} gives $E_B/E_{kp}\sim (m/M)^2$. It is clear that Eq.(\ref{sat}) is the correct saturation criterion for the magnetic field in the Weibel instabilities, valid for both electron-driven and ion-driven cases.}

\section{PIC simulations}
\label{sim}
The saturation condition Eq.(\ref{sat}) was derived for a single mode, ignoring mode-mode interactions. To study how accurately Eq. (\ref{sat}) describes the saturation level when a range of unstable modes are present, whether Eq. (\ref{ebk}) is a good estimation of the total magnetic field energy, and whether the electron gained energy is of the same order of magnitude as the magnetic field energy, we performed a series of PIC simulations with the fully explicit PIC code OSIRIS\cite{osiris}. The simulations also  provide a check on the theory of Weibel instabilities in the low growth rate regime\cite{tzouf06} and include any relativistic effects neglected in the analysis. This regime is motivated by 
turbulent astrophysical plasmas of interest around black holes which 
 are expected to have low to moderate temperature anisotropy ($\alpha\leq M/m$), moderately relativistic temperatures,  and low growth rates.
Previous PIC simulations were in the large anisotropy and large growth rate ($\alpha> M/m$ and large $\xi$) regime\cite{davidson72,msk06}. 

In practice, PIC simulations of low growth rate instabilities are more difficult  than those of a high growth rate. The number of particles used in a PIC simulation plasma is usually much smaller than that of the actual  plasma it simulates. The resulting inflated fluctuations in the simulations, including the associated collisions, set a practical lower limit on the growth rate of a collisionless 
instability that can be reliably measured.  Eqs.(\ref{gmax}) and (\ref{ebk}) show that an actual proton-electron mass ratio of $M/m=1836$ would produce too small a growth rate and too weak a magnetic field energy density to be practically measured in a simulation. We have therefore used $M/m=10-30$ in our simulations to check the ion mass scaling. Both electrons and ions were initially uniformly distributed in space, with a velocity distribution of the form $\exp(-p_x^2/p_{tx}^2)\exp(-p_y^2/p_{ty}^2)$ where $p_{x,y}$ is the normalized relativistic momentum components, $p_{x,y}\equiv (v_{x,y}/c)/\sqrt{1-v_x^2/c^2-v_y^2/c^2}$. The $z$-component of the velocity was kept to zero. 
For the electrons, $p_{tx}=p_{ty}=0.41$ were chosen, corresponding to
$\sqrt{T_e/mc^2}\simeq 0.38$ $(T_e=74$ keV) if we write $T_e/mc^2\approx p_{tx}^2/\sqrt{1+p_{tx}^2+p_{ty}^2}$.
For the ions, $p_{tx}=0.13$ and $p_{ty}=0.41$ were chosen, corresponding 
to $\sqrt{T_{Ix}/Mc^2}\simeq 0.12$,  $\sqrt{T_{Iy}/Mc^2}\simeq 0.39$, and $\alpha=9$. (For protons,  this corresponds to $T_{Ix}\simeq 15$MeV and  $T_{Iy}\simeq 145$MeV.) The simulations used 1-1/2D (one full spatial dimension, $x$, and one half-dimension, $y$, where no spacial variation was allowed, and three velocity components but with $V_z$ remaining zero throughout the runs) with a box size of 240 $c/\omega_{pe}$. 
The cell size was $\Delta x=0.1c/\omega_{pe}$ and time step was $\Delta t=0.099/\omega_{pe}$. To reduce fluctuations, 2500 to 18000 particles per cell were used for each species in typical runs. The total energy in these runs was conserved within $10^{-6}$.

\begin{figure}[!th]
\begin{center}
\includegraphics[width=6in]{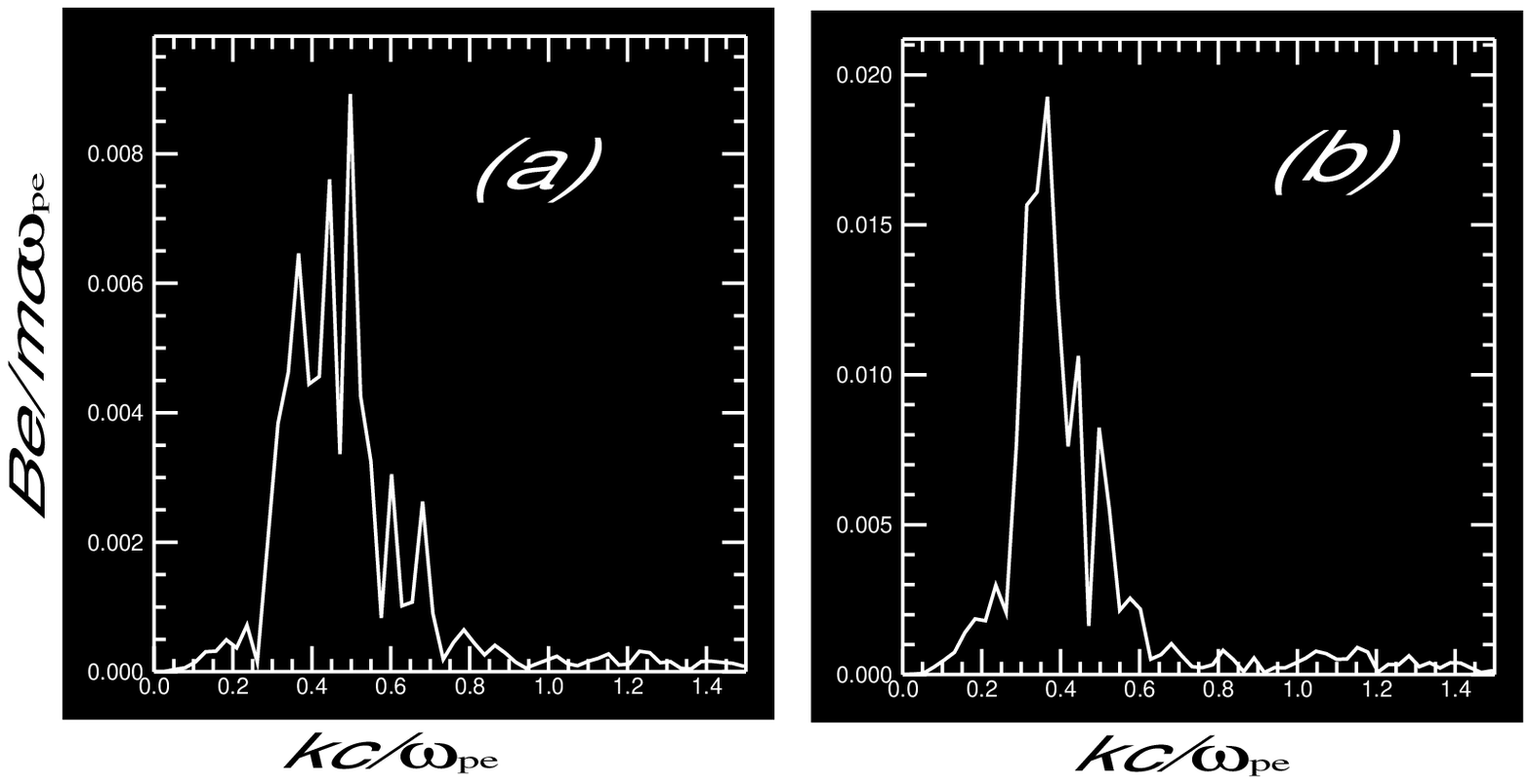}
\caption{Fourier spectrum of $B_z$ at two different times, $t_1=166\omega_{pe}^{-1}$ (a) and $t_2=305\omega_{pe}^{-1}$ (b), for the $M/m=10$ case.} 
\label{spec}
\end{center}
\end{figure}

Fig. \ref{spec} shows the Fourier spectrum of $B_z$ at two different times, $t_1=166\omega_{pe}^{-1}$ (in the linear growth stage) and $t_2=305\omega_{pe}^{-1}$ (after saturation), for the $M/m=10$ case ($\beta=3$ and 2500 particles per cell were used.)
During the linear growth stage, the growth rate $\gamma$ for the $k=0.4\omega_{pe}/c$ was measured to be $0.026\omega_{pe}$, agreeing reasonably well with the analytical result of $0.031\omega_{pe}$ obtained directly from Eq. (\ref{full_disp}). (Here, $|\xi_I|\approx0.45$ so the small $\xi_I$-expansion result of Eq. (\ref{rate}) is not accurate and the analytical result is obtained using exact values of the $Z$-function.) Following the same mode to saturation at $t=305/\omega_{pe}$, the right hand side of Eq. (\ref{sat}) was found to be $0.017\omega_{pe}$, about 65\% of the linear growth rate, consistent with Eq. (\ref{sat}) as a qualitative saturation criterion. 
This also implies  that Eq. (\ref{ebk}) 
is consistent only as an order-of-magnitude estimate.
Fig. \ref{spec} also shows the shift of the dominant mode to lower $k$ as the instability saturated, since $k^{*}$ of the maximum growth mode decreases as $\alpha$ decreases and also $k_{m}$ of the highest saturation mode is smaller than $k^{*}$.

Fig. \ref{fig-e} shows the time evolution of the 
 energy change in ions, electrons, and magnetic field ($B_z$-component), in units of the fraction of the initial ion energy. The curve for the magnetic energy is the total magnetic magnetic energy, with contributions 
integrated over the full range of $k$. 
This magnetic field energy reached  $\sim1.8\times10^{-3}$ of the initial ion energy, which is of order $1.3\times10^{-3}$ estimated from Eq. (\ref{ebk}). The electron energy change at saturation was about $3.4\times10^{-3}$, significantly higher than the gain of $1.8\times10^{-4}$ that corresponds to the gain from  intrinsic collisions in the code: The latter baseline collisional  gain was measured in a run with an isotropic ion distribution and found to depend linearly on $t$ and inverse linearly on the number of particles per cell used. Therefore, the electron energy gain in the simulation of Fig. 2 was mainly due to collisionless processes as shown.  After saturation, the electron energy remains at the same order of magnitude till the end of the simulation ($t=3000 \omega_{pe}^{-1}$ or $948\omega_{pI}^{-1}$).

The ion mass scaling of (\ref{ebk}) 
was checked by comparing the $M/m=10$ run with a run of $M/m=20$. The saturated magnetic energy should scale as $M^{-4}$ according to Eq. (\ref{ebk}), and the time to reach  saturation should scale as $M^{3/2}$ (Eq. \ref{gmax}). 
Because the ion-electron collisional energy exchange rate in our simulations was found to scale as $M^{-1}$,  as in a real plasma \cite{trub65}, it was necessary to increase the number of particles per cell to 18000 in the $M/m=20$ simulations to reduce the collisional energy exchange rate to below the collisionless instability-induced level.  Fig. \ref{M20} shows the energy changes for the $M/m=20$ case. The saturated magnetic field energy reached at $\sim1.3\times10^{-4}$ of the initial ion energy, somewhat less, but to within order of magnitude of the $5.3\times10^{-4}$ estimated from Eq. (\ref{ebk}) (using $\beta=1.5$). The electron energy change at saturation was $\sim 3.2\times10^{-4}$, roughly scaling as $M^{-3.4}$ when compared with the $M/m=10$ case. Overall, the PIC simulations support Eq. (\ref{ebk}) as an estimation of the saturated magnetic field energy and that the electron energy gain is of the same order of magnitude as the magnetic field energy.

\section{Summary and Discussions}
\label{dis}
In summary, the linear growth rate of the Weibel instability driven by the proton temperature anisotropy has been calculated in the low anisotropy limit [Eqs.(\ref{rate}) and (\ref{gmax})]. This growth rate is also used to derive an expression for the ratio of saturated magnetic field energy to the initial proton kinetic energy, $E_B/E_{kp}$, in the low anisotropy limit [Eq.(\ref{ebk})]. These results show that $E_B/E_{kp}$ scales as $(m/M)^4$ when $\sqrt{T_e/m}\sim\sqrt{T_{Ix}/M}$ and is extremely small for $\alpha=1$. 1.5D PIC simulations with a reduced ion to electron mass ratio support  the scaling relations and also show that the electron energy gain from the proton anisotropy is comparable to the gain in  magnetic field energy. Although in three dimensions one additional mode polarization is allowed and the number of unstable modes can double, we do not expect orders of magnitude change in the electron energy gain. 
From these analyses and simulations we conclude that the proton-driven Weibel instability will not  provide an efficient faster-than-Coulomb 
electron-ion temperature equilibration mechanism in an intially unmagnetized two temperature plasma with low to moderate anisotropy.

{We have also shown that the low anisotropy formulae derived here approach to the previous high anisotropy results when $\alpha\rightarrow M/m$ [Eqs.(\ref{large_rate}) and (\ref{ebk2})]. This makes explicit the implied assumption of $\alpha\geq M/m$ used in the discussion of magnetic field generation in collisionless shocks through the Weibel instability\cite{wa04,msk06}. It also provides guidance on how to properly scale PIC simulations results with reduced $M/m$. For example, PIC simulations on shocks in galaxy clusters with Mach number $M_{sh}=20$, electron thermal velocity $v_{th,e}=0.05c$, and ion thermal velocity $v_{th,e}=0.005c$ were performed using $M/m=100$ in Ref.\cite{msk06}. Since $\alpha=M_{sh}^2(m/M)(v_{th,e}/v_{th,i})^2=400>M/m$, the simulations showed an $E_B/E_{kp}\approx10^{-3}$, consistent with $m/(16M)$ predicted from Eq.(\ref{ebk2}). However, if the same $M, v_{th,e},$ and $v_{th,i}$ were used for $M/m=1836$, $\alpha=22<M/m$ and Eq.(\ref{ebk}) would predict an $E_B/E_{kp}\approx 4\times 10^{-6}$.}

Finally we wish to point out that the situation may be greatly different
for  an initially magnetized plasma 
in which electrons are confined 
to local magnetic field lines transversely.  For modes with ${\bf k}$ perpendicular to the equilibrium ${\bf B}_0$, the electrons would be  practically immobile in the direction perpendicular to both ${\bf B}_0$ and ${\bf k}$. As explained in Sec.\ref{theory}, the Weibel instability is driven by the current induced by the perturbed magnetic field $B_z$, represented by the term proportional to $\alpha$ in Eq.\ref{full_disp}. However, additional currents with a different phase are also driven by the accompanied electric field $E_y$, which are represented by the second and third terms on the left of  Eq.\ref{full_disp}. These currents are present even for an isotropic distribution and oppose mode growth. If the electrons were magnetized and could not respond to $E_y$ to produce a curent, the second term would not be present and the growth rate would be independent of $m/M$. This indicates that the required equilibrium magnetic field should be such that $\omega_{ce}>>\omega_{pI}>>\omega_{cI}$ for this to happen, where $\omega_{ce}$ and $\omega_{cI}$ are the electron and ion cyclotron frequencies in the equilibrium magnetic field. In the case of the ion-driven (electricstatic) two-stream instability with magnetized electrons, the results were indeed found to be independent of the mass ratio\cite{papa71}.
 If this also were true for the Weibel instability more generally, the saturated magnetic fields could be $E_B/E_{kp}\approx8\times10^{-3}\alpha^5/[(1+\alpha)^5(3+2\alpha)]=2.5\times10^{-4}$ for $\alpha=5.4$, significantly larger than that predicted by Eq.(\ref{ebk}).  However,  the electrons still cannot gain significant energy even in this enhanced magnetic field. 
  
\section{Acknowledgments} 
C.R.'s work was supported by the US Department
of Energy through grants No. DE-FC02-04ER54789 and DE-FG02-06ER54879.  
E.B. acknowledges support from NSF grants AST-0406799, AST
00-98442, AST-0406823, NASA grant ATP04-0000-0016, and the Aspen Center for Physics.

\newpage

\begin{figure}[!th]
\begin{center}
\includegraphics[width=6in]{renfig2.epsf}
\caption{The energy change of the ions, electrons, and magnetic field, in unit of the fraction of the initial ion energy, for the $M/m=10$ case.} 
\label{fig-e}
\end{center}
\end{figure}

\begin{figure}[!th]
\begin{center}
\includegraphics[width=6in]{renfig3.epsf}
\caption{The energy change of the ions, electrons, and magnetic field, in unit of the fraction of the initial ion energy, for the $M/m=20$ case.} 
\label{M20}
\end{center}
\end{figure}

\end{document}